\documentclass[aps,twocolumn]{revtex4}
\usepackage{graphicx}% Include figure files
\begin{document}

\newcommand{\be}{\begin{eqnarray}}
\newcommand{\ee}{\end{eqnarray}}
\newcommand{\xx}{\begin{eqnarray*}}
\newcommand{\yy}{\end{eqnarray*}}
\newcommand{\nn}{\nonumber}
\newcommand{\Vol}{{\rm Vol}}
\newcommand{\sign}{{\rm sign}}
\newcommand{\rt}[1]{c_{#1}}
\newcommand{\fa}{f_{\alpha}}
\newcommand{\fb}{f_{\beta}}
\newcommand{\Da}{\Delta_{\alpha}}
\newcommand{\Db}{\Delta_{\beta}}
\newcommand{\D}[1]{\Delta_{#1}}
\newcommand{\G}[1]{\tilde G(#1)}
\newcommand{\Sm}{M}
\newcommand{\temp}{t}
\newcommand{\mm}{\tilde m}
\newcommand{\ace}{A_{CE}}
\newcommand{\agce}{A_{GCE}}
\newcommand{\rf}[1]{${\rm Eq}.\;(\ref{#1})$}
\newcommand{\rfa}[1]{appendix \ref{#1}}
\newcommand{\rfb}[1]{$\left [{\rm Eq}.\;(\ref{#1})\right ]$}

\title{Analytical treatment of the dHvA frequency combinations due to chemical potential oscillations in an idealized
two-band Fermi liquid}
\author{Jean-Yves Fortin\footnote{ e-mail: fortin@lpt1.u-strasbg.fr}}
\affiliation{Laboratoire de Physique Th\'{e}orique (UMR CNRS 7085),
Universit\'{e} Louis Pasteur, 3 rue de l'Universit\'{e}, 67084
Strasbourg, France.}
\author{Emmanuel Perez and Alain Audouard\footnote{e-mail: audouard@lncmp.org}} \affiliation{Laboratoire
National des Champs Magn\'{e}tiques Puls\'{e}s (UMR CNRS-UPS-INSA
5147), 143 avenue de Rangueil, 31432 Toulouse, France.}
%\begin{center}
%(Received 28 November 1999)
%\end{center}
\vskip 1cm
\date{\today}
\begin{abstract}
de Haas-van Alphen oscillation spectrum is studied for an
idealized two-dimensional Fermi liquid with two parabolic bands in
the case of canonical (fixed number of quasiparticles) and grand
canonical (fixed chemical potential) ensembles. As already
reported in the literature, oscillations of the chemical potential
in magnetic field yield frequency combinations that are forbidden
in the framework of the semiclassical theory. Exact analytical
calculation of the Fourier components is derived at zero
temperature and an asymptotic expansion is given for the high
temperature and low magnetic field range. A good agreement is
obtained between analytical formulae and numerical computations.
\end{abstract}

\maketitle \vskip 2cm PACS numbers: 67.40.Vs, 67.40-w, 67.55.-s,
67.57.Jj
%\newpage

\section{Introduction}

An ongoing question regarding magnetic oscillations in
two-dimensional (2D) multiband metals deals with frequency
combinations that are observed in de Haas-van Alphen (dHvA) and
Shubnikov-de Haas (SdH) spectra even though they are forbidden
within the semiclassical theory of magnetic breakthrough (MB) by
Falicov and Stachowiak \cite{fal}. In that respect, the organic
charge transfer salts of the $\kappa$- phase constitute
experimental realizations of the linear chain of coupled orbits
introduced by Pippard \cite{pip} in the early sixties. The Fermi
surface (FS) of these compounds is composed of one closed tube
(yielding the $\alpha$-orbit in magnetic field) and two quasi
one-dimensional sheets. This FS originates from the hybridization
of one 2D tube, yielding at high magnetic field the MB-induced
$\beta$-orbit with a cross section equal to the first Brillouin
zone area. Although these compounds, in particular
$\kappa$-(ET)$_2$Cu(SCN)$_2$ (where ET stands for
bisethylenedithia-tetrathiofulvalene), have been extensively
studied, the physical origin of some of the observed combination
frequencies \cite{har96,kar96,mey} remains unclear. Actually,
besides MB-induced closed orbits which account for some
combination frequencies such as  F$_{\beta + \alpha}$ or
F$_{2\beta - \alpha}$, oscillation of the chemical potential
\cite{har96,theo-CuNCS} and MB-induced field-dependent broadening
of the Landau levels (LL) \cite{for,gvo} have been invoked in
order to interpret the "forbidden" combination frequencies such as
F$_{\beta - \alpha}$ or F$_{\beta - 2\alpha}$ observed in dHvA
data. When they are observed in SdH spectra of $\kappa$- phase
salts \cite{har96,kar96} and of other organic metals \cite{lyu02}
or layered semiconductor structures \cite{deu01} with similar FS,
these latter combination frequencies can be accounted for by
quantum interference (QI). Nevertheless, as also pointed out in
the case of organic metals whose FS constitutes a network of
coupled electron and hole orbits \cite{QI}, MB-induced LL
broadening and (or) chemical potential oscillation as well
as dimensionality certainly play
a role in the SdH oscillatory spectra.

The respective contribution of these two phenomena to the
oscillatory spectra, that certainly depend on parameters such as
the temperature, magnetic field, effective masses, scattering
rates, etc...,  needs to be quantitatively determined. In
particular, it is of primary importance to quantify the
temperature dependence of the amplitude of the various
oscillations series observed in order to check if there is a
temperature and magnetic field range in which the
Lifshits-Kosevich (LK) formula \cite{LK} still satisfactorily
applies and if addition rules such as that reported by Falicov and
Stachowiak \cite{fal} for three-dimensional FS can be derived.

In the following, a two band Fermi liquid with no MB and no
electron reservoir (although it can have a significant influence
on both the amplitude and the wave form \cite{reservoir}) is
considered \cite{theo-mu}. In such a system, only the chemical
potential oscillation have a significant contribution to the dHvA
spectrum. LK and Falicov-Stachowiak models, which assume a
constant chemical potential [grand canonical ensemble (GCE)],
predict that the dHvA Fourier spectrum contains only the
frequencies linked to each orbit and their harmonics. In this
case, the two bands are physically independent. However, for a
constant number of electrons [canonical ensemble (CE)], the system
lowers the total energy at zero temperature by sharing the
electrons between the lowest LL of the two spectra as the magnetic
field varies. In one sense, the filling of the lowest LL implies
an effective interaction between the two bands, even though MB is
not required. This leads, as already demonstrated, to a mixing of
the fundamental frequencies \cite{theo-CuNCS,theo-mu} and the
problem becomes less trivial.

\section{Model parameters}

The energy dispersion of the two bands $\alpha=0, 1$ is
characterized by band bottoms $\Delta_{\alpha}$ and effective
masses $m_{\alpha}^*$. In a magnetic field $B$, The LL of each
band  are written as:

\be\label{eq1} \epsilon_{\alpha}(n)=\Delta_{\alpha}+(h^2/2{\pi}
m_{\alpha}^*)(eB/h)(n+1/2). \ee

In zero field, the electron densities $n_{\alpha}$ of each band
satisfy the conservation equation $\sum_{\alpha}n_{\alpha}=n_e$,
where $n_e$ is the total electron density. The area of the FS
piece linked to the $\alpha$-band, $S_{\alpha}=
4\pi^2/n_{\alpha}$, is given by $S_{\alpha}=2\pi
m_{\alpha}^*/\hbar^2 (E_F-\Delta_{\alpha})$ or, equivalently,
$S_{\alpha}=eF_{\alpha}/\hbar$ where  $F_{\alpha}$ is the Fourier
frequency linked to the ${\alpha}$ orbit.

In the following, the energies are expressed in units of
$2\pi\hbar^2/m_0^*$ and we use the dimensionless variables
$x=n_e/b$, where $b=eB/h$ and $\rt{\alpha}=m_0^*/m_{\alpha}^*$ (i.
e. $c_0 = 1$ and $c_1 = m_0^*/m_1^*$). In particular we will
consider the case where $\rt{1}$ is the ratio of 2 irreducible
integers $p/q$. In the new units, the energies are expressed as
$\epsilon_{\alpha}(n)= \Delta_{\alpha}+\rt{\alpha}b(n+1/2)$. It is
also convenient to define the dimensionless fundamental
frequencies $\fa=S_{\alpha}/4\pi^2n_e$, in which case $f_0+f_1=1$.
Also defining $f_b=(\Delta_0-\Delta_1)/\rt{1} n_e$, we can express
the fundamental frequencies as a function of the effective mass
ratio $c_1$ and $f_b$:

\be\nn
f_0=\frac{\rt{1}}{1+\rt{1}}(1-f_b),\;\;{\rm and}\;\;f_1=\frac{1}{1+\rt{1}}
(1+\rt{1} f_b).
\ee

Assuming a perfect crystal, the amplitude of the Fourier
components is only driven by the thermal fluctuations through the
damping factor $R_{\alpha,p}(\temp)=p\lambda_{\alpha} /\sinh
(p\lambda_{\alpha})$, where
$\lambda_{\alpha}=2\pi^2\temp/\rt{\alpha}b$ and \emph{t} is the
temperature in unit of $2\pi\hbar^2/k_Bm_0^*$. 
 For real crystals with finite scattering relaxation time, the 
product $R_{\alpha,p}(T)R_{\alpha,p}^D$ should be considered 
instead (the Dingle damping factor is given by
$R_{\alpha,p}^D=\exp(-2\pi^2pt_D/c_{\alpha}b)$, where $t_D$ is the Dingle
temperature). 
%\textbf{\emph{(Je suppose qu'on peut
%utiliser les \'{e}quations en rempla\c{c}ant $R_{\alpha,p}$ par le
%produit $R_{\alpha,p}\times R(Dingle)$? Cela permettrait de
%traiter le cas des cristaux "r\'{e}els".)}} 
The relations between
the temperature, magnetic field and the reduced parameters are
given by:

\be\nn T(K)&=&\temp \frac{2\pi\hbar^2}{k_Ba_0^2m^*_0}=5.556\times
10^{-15} \frac{\temp}{a_0^2m_0^*},
\\ \nn
B(T)&=&b\frac{2\pi\hbar}{ea_0^2}= 4.136\times
10^{-15}\frac{b}{a_0^2}, \ee

where $a_0$ is the unit cell length. In the case of the
$\kappa$-(ET)$_2$Cu(SCN)$_2$ compound ($a_0^2=1.11\times
10^{-18}m^2, m^*_0=3.2, m^*_1=7)$, the ratio $\temp/b$ is equal to
2.38$\times$T(K)/B(T).

The chemical potential is solution of an implicit equation which
cannot be solved analytically, except at zero temperature (see
Section IV). Nevertheless, assuming $t/b$ is high enough for the
chemical potential oscillations are weak, an asymptotic expansion
can be derived at the lowest order. The deduced expressions allow
for the calculation of the Fourier components of the dHvA spectrum
as reported in  Section III. In the LK and FS theories their
amplitude, which is governed by the thermal damping factor,
exponentially decreases as the $t/b$ ratio, the harmonics' order
and the effective masses increase. Although such feature is also
observed in most cases in the CE, it is demonstrated hereafter
that the factor in front of the thermal damping factor is
generally not the same as in the LK theory. This may lead to a non
monotonous temperature dependence of the amplitudes. E. g., the
Fourier amplitude of the harmonics may even vanish at a finite
temperature depending on the effective masses and the harmonic
order. This behavior is related to the complex oscillations of the
chemical potential that implie fluctuations of the area, hence frequencies,
of the electronic trajectories defining the FS. 
More precisely, these frequencies  are proportional
to the chemical potential $\mu-\Delta_{\alpha}$ and therefore
oscillate. It is then necessary to define the problem in terms
of zero field frequencies, the oscillations of the chemical 
potential modifying only the amplitudes.
%\emph{Je suppose que tu as une bonne raison
%pour parler de cela mais je ne vois pas laquelle? Par ailleurs,
%Champel (Phys. Rev. B \textbf{69}) fait remarquer que dans le CE,
%les f$_{\alpha}$ ne sont pas de vraies fr\'{e}quences puisqu'elles
%oscillent avec le champ mais tu dis a peu pr\`{e}s la m\^{e}me
%chose juste avant l'\'{e}quation (5), non?}.

The analytical formulae deduced in sections III and IV are
compared to the Fourier spectra deduced from numerical computations of
the chemical potential both at zero and finite $t/b$. Discrete
Fourier transforms are calculated with a Blackman window in a
given field range from $b_{min}$ to $b_{max}$. The absolute value
of the Fourier amplitude ($A$) at a given frequency for a mean
field value $b_{mean}=2/(1/b_{min}+1/b_{max})$ is determined from
the amplitude of the discrete Fourier transform ($A_{calc}$) as $A
= 4A_{calc}/0.84(1/b_{min}-1/b_{max})$. As reported hereafter, an
excellent agreement between analytical expressions and numerical
computations is obtained at zero temperature while the asymptotic
expansion is a good approximation in the large $\temp/b$ range.

\section{Fourier components of the magnetization
in the high temperature regime}

To evaluate the thermodynamical quantities, we begin by
considering the Fourier transform of the grand potential $\Omega$
in the GCE for the two band system. It is expressed as:

\be\label{eq2}
\Omega&=&\sum_{\alpha}\Omega_{\alpha}
\\
\nn
\Omega_{\alpha}&=&-\frac{1}{2\rt{\alpha}}(\mu-\Delta_{\alpha})^2+
\frac{b^2\rt{\alpha}}{2}\left [
\frac{1}{12} \right .
\\ \nn
&+&\left .\sum_{p\ge 1}\frac{(-1)^p}{\pi^2p^2}R_{\alpha,p}(\temp)
\cos\left (2\pi p
\frac{\mu-\Delta_{\alpha}}{\rt{\alpha}b}\right )\right ],
\ee

The Fourier transform of the magnetization $M=-\partial
\Omega/\partial B$ at fixed chemical potential $\mu$ contains
cosine functions of the arguments $2\pi p F_{\alpha}/B=2\pi p \fa
x$ with $p$ integer, and therefore the frequencies involved in the
spectrum are the individual frequencies $F_{\alpha}$ plus their
harmonics. This is the LK result. In the GCE, there is therefore
no mixing of the form $kf_0+lf_1$ with $(k,l)$ positive or
negative integers. In the CE, the magnetization is expressed as
$M=-\partial F/\partial B$, where $F=\Omega+n_e\mu$ is the free
energy. Contrary to the GCE case, it is known that the Fourier
spectrum of the magnetization contains such  $kf_0+lf_1$
combinations, where the $\fa=(\mu_0-\Delta_{\alpha})/
c_{\alpha}n_e$ are defined relatively to the zero field chemical
potential $\mu_0=\mu(B=0)$. The chemical potential is solution of
the equation:

\be\label{eq3}
& &\sum_{\alpha}\frac{\mu-\Delta_{\alpha}}{\rt{\alpha}}
=n_e
\\ \nn
& &-b\sum_{\alpha}\sum_{p\ge 1}\frac{(-1)^p}{\pi
p}R_{\alpha,p}(\temp) \sin \left ( 2\pi p
\frac{\mu-\Delta_{\alpha}}{\rt{\alpha}b}\right ).
\ee

At zero field, the last term vanishes, and we get the simple
relation $\mu(B=0)=\mu_0=\Delta_{\alpha}+\rt{\alpha}n_{\alpha}$.
At finite magnetic field, the chemical potential oscillates, and
is solution of the self-consistent equation \rf{eq3}. If we
replace $\mu$ in \rf{eq2} by this solution, the free energy is
obtained using the relation $F=\Omega+n_e\mu$. It can be noticed
that the series in \rf{eq3} is proportional to the small parameter
$b$. We may therefore replace $\mu$ in the cosine functions by its
zero magnetic field value $\mu_{0}$:

\be\label{eq4}\nn
\sum_{\alpha}\frac{\mu-\Delta_{\alpha}}{\rt{\alpha}} \approx n_e
-b\sum_{\alpha}\sum_{p\ge 1}\frac{(-1)^p}{\pi
p}R_{\alpha,p}(\temp) \sin \left ( 2\pi p \fa x\right ).
\ee

This relation can be expressed as:

\be\label{eq5}
\mu\approx\mu_0-b\frac{\sum_{\beta}M_{\beta}}{\sum_{\beta}\rt{\beta}^{-1}},
\ee

with $M_{\beta}=\sum_{p\ge 1}(-1)^pR_{\beta,p}(\temp)/\pi
p\times\sin (2\pi p\fb x)$. Replacing $\mu$ in \rf{eq2} by this
expression, and then computing $F=\Omega+n_e\mu$, we obtain a
finite temperature approximation for the free energy. The
approximation \rf{eq6} shows that the dominant oscillatory
behavior of the potential leads to  frequency combinations.
Indeed, in the oscillatory part of \rf{eq2}, the arguments of the
cosine functions are themselves oscillating:

\be\label{eq6}
\cos\left (2\pi p\frac{\mu-\Delta_{\alpha}}{\rt{\alpha}b}\right )
&=&\Re \left \{
\exp (2i\pi p\fa x)\right .
\\ \nn
&\times &\left .\prod_{\beta}\exp (-2i\pi pw_{\alpha}M_{\beta}(x))
\right \},
\ee

where $w_{\alpha}=\rt{\alpha}^{-1}/\sum_{\beta}\rt{\beta}^{-1}$
are mass weights. The functions $\exp(-2i\pi
pw_{\alpha}M_{\beta}(x))$ are periodic functions of $x$ with $\fb$
period, therefore they can be expanded in their Fourier series:

\be
\exp (-2i\pi pw_{\alpha}M_{\beta})=\sum_{q=-\infty}^{\infty}
B_{\alpha,\beta}(p,q)\exp(2i\pi q f_{\beta} x).
\ee

We obtain a product of several Fourier series, and expanding the
product and combining the oscillating functions lead to the
combinations between the $\fa$ frequencies.

If we try to go beyond this approximation, the scheme breaks down.
Indeed, $\delta\mu=\mu-\mu_0$ is a small parameter, and we would
like to expand the sine functions in \rf{eq3} relatively to this
parameter. The first order is the case we discussed before, and at
the next order we obtain

\be\label{eq7.1}
\sin\left (
2\pi p \frac{\mu-\Delta_{\alpha}}{\rt{\alpha}b}\right )
&\approx& \sin\left (2\pi p \fa x\right )
\\ \nn &+&\frac{2\pi p}{\rt{\alpha}b}
\cos\left (2\pi p\fa x\right )\delta\mu.
\ee

The last term in \rf{eq7.1} is proportional to
$p\,\delta\mu/b\simeq p$ which is not small. It is therefore
difficult to expand \rf{eq4} as a series of $\delta\mu$. However,
the first order approximation works for high  $\temp/b$ values
(this is checked hereafter by solving numerically the
self-consistent relation \rf{eq3}), since the oscillations of the
chemical potential are negligible in this range. The oscillating
part of the magnetization is defined in units of $\hbar e/m_0^*$
by $m_{osc}\propto (n_e/b^2)\partial F/\partial x$ or
$m_{osc}=\sum_{F}\ace(F)\sin(2\pi Fx)$. The dominant value is:

\be\nn
m_{osc}&\simeq&-\frac{n_e}{\sum_{\alpha}c_{\alpha}^{-1}}\sum_{\alpha,\beta}M_{\alpha}(x)
M_{\beta}'(x)
\\ \nn
&+&\sum_{\alpha}\frac{n_ec_{\alpha}}{2\pi^2}\sum_{p_{\alpha}\ge 1}
\frac{(-1)^{p_{\alpha}}}{p_{\alpha}^2}R_{\alpha,p_{\alpha}}
\frac{\partial}{\partial x}\Re \left [ \exp(2i\pi p_{\alpha}
f_{\alpha} x)\right .
\\ \nn
&\times&\left .\sum_{q_0}B_{\alpha,0}(p_{\alpha},q_0)
\exp(2i\pi q_0 f_0 x) \right .
\\ \label{eq7.2}
&\times& \left .
\sum_{q_1}B_{\alpha,1}(p_{\alpha},q_1)\exp(2i\pi q_1 f_1 x)\right ].
\ee

The various terms in \rf{eq7.2} comes from the derivation of
$n_e\mu +\Omega$ using \rf{eq2} and \rf{eq5}. In the large
$\temp/b$ range, the coefficients $B_{\alpha,\beta}$ can be
estimated by noticing that $M_{\beta}$ is dominated by the first
term of its Fourier series $M_{\beta}(x)\simeq
-R_{\beta,1}/\pi\sin(2\pi f_{\beta}x)$. In this case those
coefficients are simply related to Bessel functions:

\be
B_{\alpha,\beta}(p,q)\simeq J_q(2pw_{\alpha}R_{\beta,1}).
\ee

The arguments of the Bessel functions are rather small since
$R_{\beta,1}\simeq 2\lambda_{\beta}\exp(-\lambda_{\beta})\ll 1$.
The amplitudes $\ace(F)$ and the related masses can be extracted
within this approximation at the lowest order, and for different
frequency combinations. The $m_{\alpha}^{th}$ harmonics of
$f_{\alpha}$, result from two contributions: one coming from the
product of $M_{\alpha}(x)M_{\alpha}'(x)$ in \rf{eq7.2}, and the
other from the solutions $(p_{\alpha}+q_{\alpha}=\pm m_{\alpha},
q_{\beta\neq \alpha}=0)$ and $(p_{\beta}+q_{\beta}=0,
q_{\alpha}=\pm m_{\alpha})$ in the second term of the right hand
side of \rf{eq7.2}. For $m_{\alpha}=1$, we simply obtain:

\be\nn
\ace(f_{\alpha})&\simeq& \frac{n_ec_{\alpha}f_{\alpha}}\pi R_{\alpha,1}
\\ \label{eq7.3}
&\simeq& 4\pi n_ef_{\alpha}\frac{\temp}{b}\exp(-2\pi^2\frac{\temp}
{c_{\alpha}b}),
\ee

which is the same amplitude as in the GCE. Indeed the main contribution
comes from the terms $B_{\alpha,\alpha}(1,0)B_{\alpha,\beta}(1,0)\simeq 1$,
the next corrective terms are smaller and correspond to the third order in $R_{\alpha,1}$.
The amplitudes of the second harmonics $\ace(2f_{\alpha})$ are different
in the CE:

\be\nn
\ace(2f_{\alpha})&\simeq&\agce(2f_{\alpha})
\\ \label{eq7.4}
&\times&\left (1-4\pi^2\frac{\temp}{c_{\alpha}b}w_{\alpha}\right )
\\ \nn
\agce(2f_{\alpha})&\simeq&-4\pi
n_ef_{\alpha}\frac{\temp}{b}\exp(-4\pi^2
\frac{\temp}{c_{\alpha}b}). \ee

Within this approximation, $\ace(2\fa)$ vanishes and changes sign
at $\temp/b=c_{\alpha}/(4\pi^2w_{\alpha})$, due to the additional
factor in \rf{eq7.4}. Additional corrective terms may modify this
value, but a more explicit expansion would be needed in order to
get a better accuracy.

The amplitudes for the sum and difference of $f_0$ and $f_1$ are
computed in the same way and we obtain:

\be \label{eq7.5}
\ace(f_0\pm f_1)\simeq\frac{n_e}{\sum_{\alpha}c_{\alpha}^{-1}}
\frac{f_1\pm f_0}{\pi}R_{0,1}R_{1,1}
\\ \nn
\simeq \frac{16\pi^3n_e}{\sum_{\alpha}c_{\alpha}^{-1}}
\frac{f_1\pm f_0}{c_0c_1}\frac{\temp^2}{b^2}
\exp(-2\pi^2(\frac{1}{c_0}+\frac{1}{c_1})\frac{\temp}{b}).
\ee

The amplitudes for $\fa+2\fb$, $\alpha\neq\beta$, are

\be\nn
\ace(\fa+2\fb)\simeq\frac{n_e(\fa+2\fb)}{\pi\sum_{\alpha}c_{\alpha}^{-1}}
R_{\alpha,1}\times
\\ \label{eq7.6}
\left (\frac{R_{\beta,2}}{2}+(\frac{w_{\alpha}}{2}+w_{\beta})
R_{\beta,1}^2\right )
\ee

leading to:

\be \frac{\ace(\fa+2\fb)}{\ace(\fb+2\fa)}\simeq
\frac{\fa+2\fb}{\fb+2\fa}\frac{1+(1+w_{\beta})\lambda_{\beta}}{
1+(1+w_{\alpha})\lambda_{\alpha}}. \ee

The amplitudes given by Eqs. \ref{eq7.3} to \ref{eq7.6} are
generally the largest that can be extracted from data analysis.
The ratio

\be
\frac{\ace(\fa-\fb)}{\ace(\fa+\fb)}\simeq \frac{\fb-\fa}{\fa+\fb},
\ee

is constant at high temperature indicating that these two
oscillations have the same mass
$m^*(\fa\pm\fb)=m^*_{\alpha}+m^*_{\beta}$ that can be extracted by
plotting the logarithm of $(b/\temp)^2\ace(\fa\pm\fb)$ versus
$\temp/b$, which should be a straight line in this limit. More
generally the addition rule

\be \label{7.7} m^*(k\fa+l\fb)=|k|m^*_{\alpha}+|l|m^*_{\beta} \ee

can be derived. Nevertheless, it should be kept in mind that the
$\temp/b$ dependance of the prefactor is not simple. Therefore,
each case has to be treated separately. E. g., in the high $t/b$
range, Eq. \ref{eq7.6} can be rewritten as:

\be\nn  \label{7.8}
\ace(\fa+2\fb)\simeq\frac{4n_e(\fa+2\fb)\lambda_{\alpha}\lambda_{\beta}}{\pi\sum_{\alpha}c_{\alpha}^{-1}}\\
\times[1+2\lambda_{\beta}(1+
w_{\beta})]\exp[-(\lambda_{\alpha}+2\lambda_{\beta})].
 \ee

This leads to $m^*(\fa+2\fb)=m^*_{\alpha}+2m^*_{\beta}$, as
predicted by Eq. \ref{7.7}, although this effective mass is
derived in the high temperature range by plotting
$\log[A_{CE}(\fa+2\fb)(b/t)^3]$ vs.
%$\log[A_{CE}(\fa+2\fb)(b/t)^2/[1+(4\pi^2/c_{\beta})
%(1+w_{\beta})(t/b)]]$ vs.
$t/b$. 
%\textbf{\emph{ (tu peux v\'{e}rifier que je ne me suis pas
%tromp\'{e}?)}}.  
%For high temperature, the asymptotic factor in front of the dominant
%exponential damping is proportional to $(\temp/b)^3$.
Oppositely, the asymptotic factor in front of the dominant exponential
damping is always proportional to $\temp/b$ in the GCE.

\begin{figure}                                               % Fig 1
\centering
\resizebox{\columnwidth}{!}{\includegraphics*{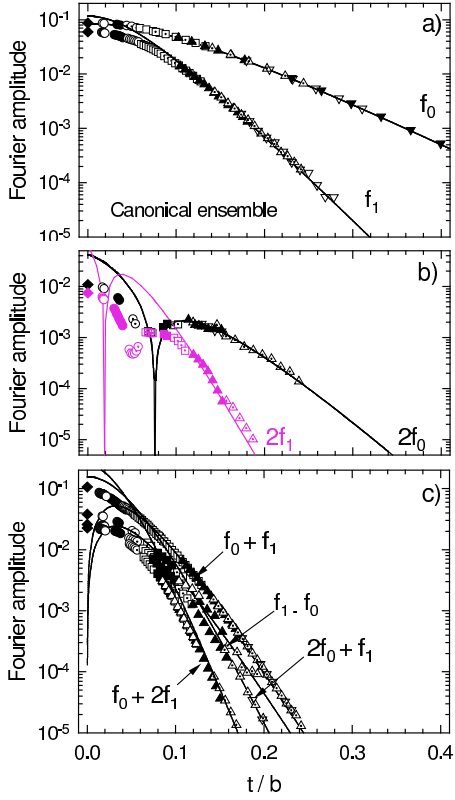}}
\caption{\label{A(tx)_CE} Reduced temperature ($t$) and magnetic
field ($b$) dependence of the dHvA oscillations amplitude for the
canonical ensemble. The band parameters are $\Delta_0 - \Delta_1$
= 1/10, c$_0^{-1}$ = 1, c$_1^{-1}$ = 2 (leading to f$_0$ = 4/15
and f$_1$ = 11/15). Symbols are data deduced from numerical
computations in the reduced temperature range from t = 0 to t =
0.009. Solid lines are deduced from the high $t/b$ approximation
given by Eqs. (\ref{eq7.3}) to (\ref{eq7.6}). Note that,
\rf{eq7.4} predicts that $A(2f_0)$ and $A(2f_1)$ vanish at
$\temp/b=c_{\alpha}/(4\pi^2w_{\alpha})$.}
\end{figure}

\begin{figure}                                               % Fig 2
\centering
\resizebox{\columnwidth}{!}{\includegraphics*{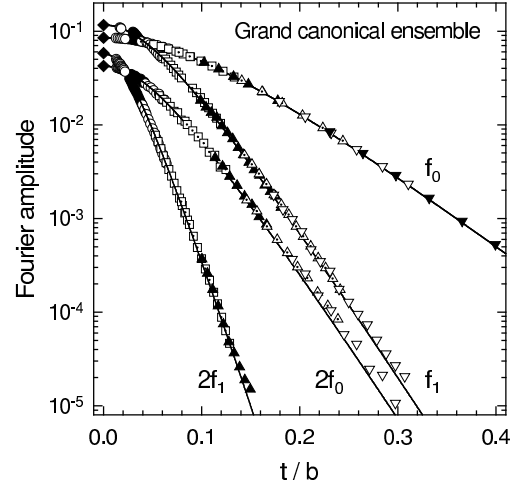}}
\caption{\label{A(tx)_GCE} Same as Fig.\ref{A(tx)_CE} for the
grand canonical ensemble. Solid lines are deduced from Eqs.
(\ref{eq7.3}) and (\ref{eq7.4}). }
\end{figure}

\begin{figure}                                               % Fig 3
\centering
\resizebox{\columnwidth}{!}{\includegraphics*{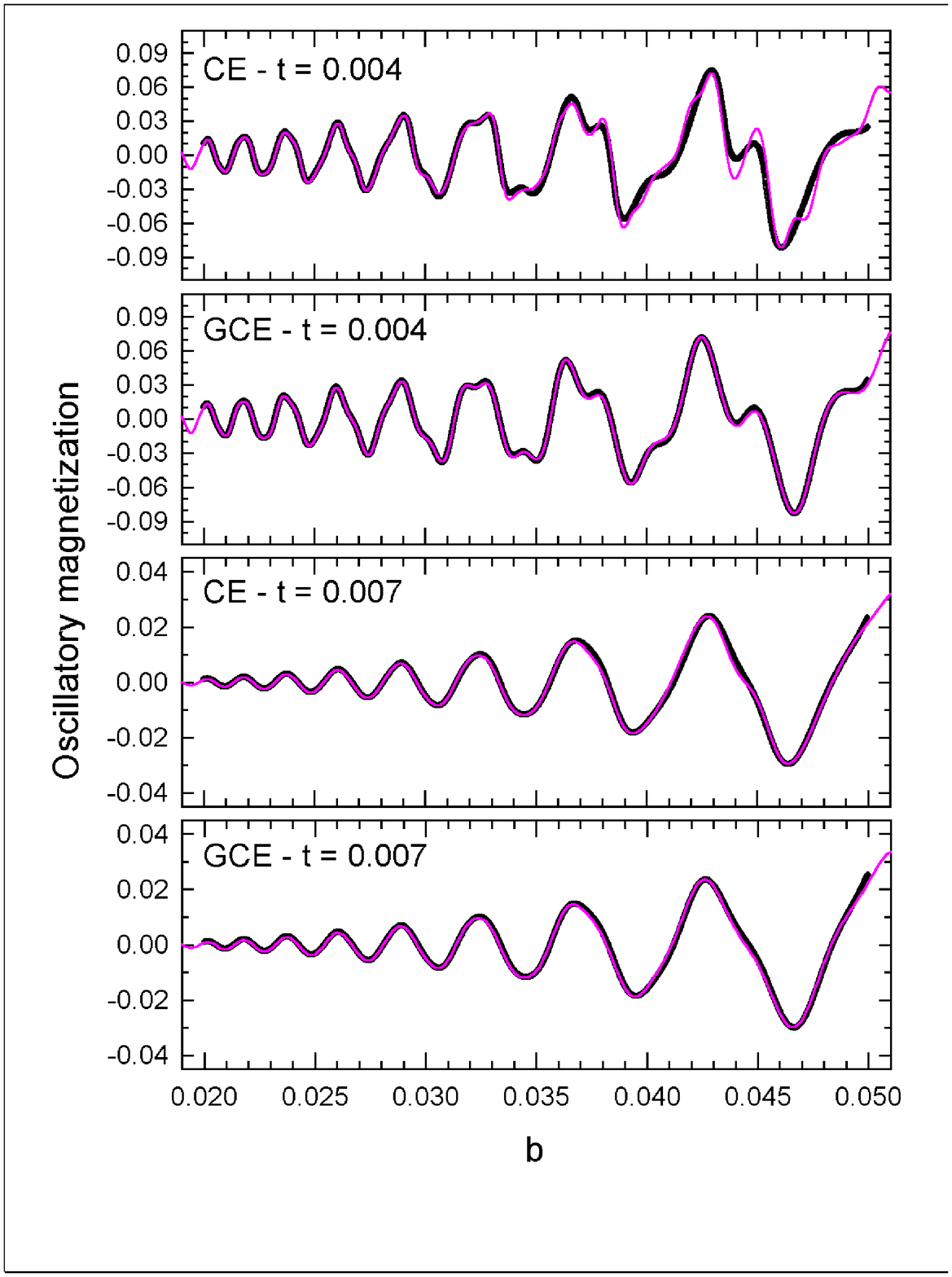}}
\caption{\label{Moscill} DHvA oscillations deduced from numerical
computations for the canonical (CE) and grand canonical (GCE)
ensembles, at two different temperatures (t = 0.004 and t =
0.007). The band parameters are the same as in Fig.
\ref{A(tx)_CE}. Thin lines are deduced from Eqs. (\ref{eq7.3}) to
(\ref{eq7.6}).}
\end{figure}

The asymptotic expressions \rf{eq7.3} to \rf{eq7.6} (see the solid
lines in  Figs. \ref{A(tx)_CE} and \ref{A(tx)_GCE}) are compared
to the amplitudes of the relevant Fourier components deduced from
the direct numerical resolution of the chemical potential given by
\rf{eq3} (symbols in  Figs. \ref{A(tx)_CE} and \ref{A(tx)_GCE}).
First, the non monotonous behavior of $A_{CE}(2f_{\alpha})$
predicted by  \rf{eq7.4} can be observed in numerical data of Fig.
\ref{A(tx)_CE}b although at a shifted $t/b$ value. More generally,
a good agreement between numerical and analytical data is obtained
for $t/b\gtrsim0.05$ and $t/b\gtrsim0.1$ ($t/b\gtrsim0.15$ for
$f_0+2f_1$ and $2f_0+f_1$) in the GCE and the CE, respectively.
This is also evidenced in Fig. \ref{Moscill} in which the
magnetization oscillations deduced from numerical resolution of
\rf{eq3} at $t=4\times 10^{-3}$ and $t=7\times 10^{-3}$ are
compared to the analytical approximations given above, in the
field range $0.02\leq b \leq 0.05$.

\section{Zero temperature Fourier analysis}

At low temperature, the amplitudes differ strongly from the LK theory,
due to the importance of the chemical potential fluctuations.
At zero temperature, the chemical potential equation \rf{eq3}
can be solved exactly and expressions for the total energy and magnetization
are obtained.
The total energy $E$ is a function of $x$ and can be
expressed as

\be\label{eq8}
E(x)=E_0+\frac{b^2}{2}G(x),
\ee

where $G(x)$ is an oscillatory function, and $E_0$ is the energy
in absence of field. The latter is given by

\be\nn
E_0&=&\frac{\rt{1}}{2(1+\rt{1})}n_e^2+\frac{n_e}{1+\rt{1}}
(\rt{1}\D{0}+\D{1})
\\ \label{eq9}
&-&\frac{(\D{0}-\D{1})^2}{2(1+\rt{1})}.
\ee

In the following, we compute $G$ and its Fourier transform by
using a combinatorial analysis. We give here the main analytical
arguments to obtain the final result. If we suppose that $\D{0}$
is large compared to $\D{1}$ (the final result does not depend on
this assumption), we first begin to fill the LL corresponding to
the band $\alpha=1$ since it has a lower band bottom. If $n_1(x)$
is the number of LL we have to fill before reaching the first
level of the band $\alpha=0$, then there are $b(x-n_1(x))=b\times
y$ electrons left we have to share between the two bands. We
obtain $n_1(x)=[(\D{0}-\D{1})/\rt{1}b-1/2]$, where $[\ldots]$
means the integer part. We can relabel in this case the LL
energies of the 1-band as $\epsilon_{1}(n)=
\tilde\D{1}+\rt{1}b(n+1/2)$, $n\ge 0$, with
$\tilde\D{1}=\D{1}+c_1b[ (\D{0}-\D{1})/\rt{1}b-1/2]$ and
$\tilde\D{1}\rightarrow\D{0}$ when $b\rightarrow 0$. The LL
corresponding to $n<0$ of the 1-band are already filled with
$bn_1(x)$ electrons and there are now $b\times y$ electrons for
the remaining LL in the bands 0 and 1. If we assume that
$\rt{1}=p/q$ is a rational with $(p,q)$ positive integers, then we
can truncate the total Landau spectrum, made of the superposition
of the LL of the two bands, by noticing that this spectrum has a
periodicity equal to $p+q$. We call periodicity the minimum number
of LL after which the relative positions of the LL between the
bands 1 and 2 repeat themselves. We can therefore decompose $x$ as
$x=n_1(x)+k_0(p+q)+ m_0+m_1+r$, with $k_0$, $m_0$ and $m_1$
integers. $k_0$ is the maximum number of blocks of $p+q$ LL
completely filled, $m_{\alpha}$ is the number of LL in the
$\alpha$-band filled, and $0\le r<1$ is the fraction of the last
LL occupied by the remaining $b\times r$ electrons (either in band
0 or 1 in accordance with the constraint that the total energy
should be minimum). We have the relations $0\le m_0 \le p$ and
$0\le m_1 \le q$, $0\le m_0+m_1<p+q$. $k_0$ is simply defined by
$k_0=[y/(p+q)]$, $r=x-[x]$ and the sum $m_0+m_1=[y-k_0(p+q)]\equiv
\Sm$. We finally have to find either $m_0$ or $m_1$, since the sum
is determined, and the last LL occupied by the $b\times r$
electrons. The total energy is

\be\label{eq10}
E&=&b\sum_{n=0}^{n_1}(\D{1}+\rt{1}b(n+1/2))
\\ \nn
&+&b\sum_{n=0}^{k_0p+m_0-1}(
\D{0}+b(n+1/2))
\\ \nn
&+&b\sum_{n=0}^{k_0q+m_1-1}(\tilde\D{1}+\rt{1}b(n+1/2))
\\ \nn
&+&b\times r{\rm min}(\epsilon_0(k_0p+m_0),\epsilon_1(k_0q+m_1)).
\ee

The first term gives the bottom energy
$E_b=(\tilde\D{1}^2-\D{1}^2)/2\rt{1}$ (the energy corresponding to
the LL of the 1-band filled between $\D{1}$ and $\D{0}$). We
obtain $m_0$ or $m_1$ by noticing that they are independent of
$r$. Therefore, we compute $E$ as a function of, e. g., $m_0$ for
$r=0$, and minimize the energy with respect to $m_0$. We obtain a
quadratic function of $m_0$, and the minimum is given by the
integer closest to the bottom of the parabolic curve:

\be\label{eq11}
m_0(y)=
\left  [\frac{\rt{1}}{1+\rt{1}}[y]+\frac{\tilde\D{1}-\D{0}}{b(1+\rt{1})}+
\frac{1}{2}\right ].
\ee

The energy can be put as $E=E_b+b^2E_c(k_0,\Sm,r)/2$. The effective
inverse field $x=n_e/b$ (or $y$) can be varied by either changing the 
magnetic field $b$ or changing the density of electrons $n_e$. In the latter
case, all quantity depending only on $b$ is constant, such as $E_b$
for example. This is especially useful if we want to extract the
periodical part of $E$ on the top LL. The periodicity of the energy 
is directly related to the
periodicity of the LL, and this happens every time that $(p+q)b$
electrons are introduced in the system, hence at fixed magnetic
field $b$. Following this argument, the $k_0$ dependence in $E_c$
can be removed by iteration, if we express $E_c(k_0)$ as function
of $E_c(k_0-1)$. We obtain

\be\label{eq12}
& &E_c(k_0,\Sm,r)=E_c(0,\Sm,r)
\\ \nn
&+&\frac{\rt{1}}{1+\rt{1}}(y^2-(\Sm+r)^2)
+2\frac{y-\Sm-r}{b(1+\rt{1})}
(\rt{1}\D{0}-\tilde\D{1}).
\ee

We then separate the terms depending on $\Sm$ and $r$ and the terms
depending on $y$:

\be \nn
E&=&E_b+\frac{b^2y^2}{2}\frac{\rt{1}}{1+\rt{1}}
+\frac{yb}{1+\rt{1}}(\rt{1}\D{0}+\tilde\D{1})
\\ \nn
&+&\frac{b^2}{2}\left \{E_c(\Sm,r)-\frac{\rt{1}}{1+\rt{1}}(\Sm+r)^2\right .
\\ \label{eq13}
&-&\left .\frac{2(\Sm+r)}{b(1+\rt{1})}(\rt{1}\D{0}+\tilde\D{1})\right \}.
\ee

We can check that the sum of the first 3 terms on the right hand
side of the equation \rf{eq13} are equal to $E_0$ [see \rf{eq9}]
with some additional terms:

\be\nn
E_0&=&E_b+\frac{b^2y^2}{2}\frac{\rt{1}}{1+\rt{1}}
+\frac{yb}{1+\rt{1}}(\rt{1}\D{0}+\tilde\D{1})
\\ \label{eq14}
&-&\frac{1}{2(1+\rt{1})}\left (\tilde\D{1}-\D{0}\right )^2.
\ee

We can therefore identify the $G$ function in \rf{eq8} as a periodic
function of the variable $y$:

\be \label{eq15}
G(y)&=&m_0(y)^2+\rt{1}\left ([y]-m_0(y)\right )^2
\\ \nn
&+&\rt{1}\delta (
[y]-2m_0(y))+(y-[y])\left \{
m_0(y)+\frac{1}{2}+ \right .
\\ \nn
& &\rt{1}([y]-m_0(y)+\frac{1}{2})
\\ \nn
&-&\left .\left |\rt{1}\delta+\rt{1}([y]-m_0(y)+\frac{1}{2})-m_0(y)-\frac{1}{2}
\right |\right \}
\\ \nn
&-&\frac{\rt{1}}{1+\rt{1}}y^2-\frac{\rt{1}(1-\rt{1})}{1+\rt{1}}\delta y
+\frac{\rt{1}^2}{1+\rt{1}}\delta^2,
\ee

where $\delta=(\tilde\D{1}-\D{0})/\rt{1}b$, $y$ being constrained
to be inside the interval $0\le y<(p+q)$. Indeed, using the fact
that $m_0(y+p+q)=m_0(y)+p$, we check that \rf{eq15} satisfies
$G(y+p+q)=G(y)$. In the following, we expand $G(y)$ in Fourier
series, and, using $y=x-n_1(x)$, express the result as a function
of $x$, in order to obtain the relation \rf{eq8}. The Fourier
coefficients are defined by:

\be\nn
\tilde
G_l&=&\int_{-(p+q)/2}^{(p+q)/2}G(y)\exp\left (-2i\pi\frac{ly}{p+q}\right )dy
\\ \label{eq16}
G(y)&=&\frac{1}{p+q}\sum_{l=-\infty}^{\infty}\tilde G_l
\exp\left (2i\pi\frac{ly}{p+q}\right ).
\ee

Defining the frequencies $\omega=2\pi l/(p+q)$, the integration
over $y$ can be split into small parts (assuming for example that
$(p+q)/2$ is integer):

\be\label{eq17}
\tilde G=\sum_{l=-(p+q)/2}^{(p+q)/2}\exp(-i\omega l)
\int_0^1drG(l+r)\exp(-i\omega r).
\ee

In \rf{eq15}, $m_0(l+r)=m_0(l)$, and $G$ is a second order polynomial function
in the variable $r$, hence the integration over $r$ is straightforward since
it uses the integrals $\int_0^1\exp(-i\omega r)r^kdr$, with
$k=0,1,2$.
In the limit of irrationnal ratio $\rt{1}$, as $p+q\rightarrow \infty$,
the computation is simpler and we assume that the physical quantities
should be smooth functions of $\rt{1}$. In this limit:

\be \nn
\G{\omega}&=&\int_{-\infty}^{\infty}G(y)\exp\left (-i\omega y\right )dy
\\ \label{eq18}
G(y)&=&\int_{-\infty}^{\infty}\G{\omega}
\exp\left (i\omega y\right )d\omega.
\ee

In \rf{eq17} the sum is now from $-\infty$ to $\infty$ and we have
to evaluate the following distributions

\be\nn
& &\sum_{l=-\infty}^{\infty}\exp(-i\omega l),
\sum_{l=-\infty}^{\infty}\exp(-i\omega l)l,
\sum_{l=-\infty}^{\infty}\exp(-i\omega l)l^2,
\\ \nn
& &\sum_{l=-\infty}^{\infty}\exp(-i\omega l)m_0(l),
\sum_{l=-\infty}^{\infty}\exp(-i\omega l)m_0^2(l),
\\ \nn
& &
\sum_{l=-\infty}^{\infty}\exp(-i\omega l)m_0(l)l
\ee

in order to obtain the final amplitudes
$\G{\omega}$. The simplest series is

\be\label{eq19}
\sum_{l=-\infty}^{\infty}\exp\left (-i\omega l\right )
=2\pi\sum_{l=-\infty}^{\infty}\delta (\omega-2\pi l)\equiv S(\omega).
\ee

The other series $\sum_{l=-\infty}^{\infty}\exp(-i\omega l)l^k$ with $k=1,2$
are derivatives of $S(\omega)$, and include therefore derivatives of
$\delta$ functions. It appears that these derivatives cancel each other
in the final result. The remaining distributions can also be written in term
of $S(\omega)$ function and its derivatives:

\be\label{eq20}
& &\sum_{l=-\infty}^{\infty}\exp\left (-i\omega l\right )m_0(l)\propto
\frac{\rt{1}\delta}{1+\rt{1}}S(\omega)
\\ \nn
&+&\sum_{l\ne 0}\frac{(-1)^l\exp(2i\pi l\rt{1}\delta/(1+\rt{1}))}
{2i\pi l}S\left (\omega-\frac{2\pi l\rt{1}}{1+\rt{1}}\right ),
\ee

and

\be\nn
\sum_{l=-\infty}^{\infty}\exp\left (-i\omega l\right )m_0(l)^2\propto
\left (\frac{1}{12}+\frac{\rt{1}^2\delta^2}{(1+\rt{1})^2}\right )S(\omega)
\\ \nn
+\sum_{l\ne 0}
\frac{(-1)^l\exp(2i\pi l\rt{1}\delta/(1+\rt{1}))}{i\pi l}
\times
\\ \label{eq21}
\left (-\frac{1}{2i\pi l}+\frac{\rt{1}\delta}
{(1+\rt{1})}\right )
S\left (\omega-\frac{2\pi l\rt{1}}{1+\rt{1}}\right ).
\ee

Finally, the series $\sum_{l=-\infty}^{\infty}\exp(-i\omega
l)m_0(l)l$ depends only on the derivatives of $S$. The details for
obtaining \rf{eq20} and \rf{eq21} are given in \rfa{app.1}.
$\tilde G(\omega)$ can be expressed as a sum of delta functions at
frequencies corresponding to the arguments of the function $S$.
After some tedious computations, we arrive at a Fourier spectrum
containing frequency combinations $l_0f_0+l_1f_1$. In order to obtain
$G(x)$ as a Fourier series, we have to replace $y$ by $x-n_0(x)$
in the cosine functions and express the result as a function of
$x$ only. Since $n_0(x)$ is integer, we always have $\cos(2\pi
ly)= \cos(2\pi lx)$. It is also useful to notice that
$y+\delta=x(1-f_b)$. Finally we obtain

\be\label{eq22}
& &G(x)=\frac{1}{12}\frac{1+4\rt{1}+\rt{1}^2}{1+\rt{1}}
\\ \nn
%\label{eq23}
&-&\frac{\rt{1}}{1+\rt{1}}\sum_{l=1}^{\infty}
\frac{1}{\pi^2l^2}\cos\left (2\pi l(f_0+f_1)x\right )
\\ \nn
%\label{eq25}
&+&\sum_{l\ge 1}\frac{(-1)^{l}}
{2\pi^4l^4}\frac{(1+\rt{1})^3}{\rt{1}^2}
\left (1-\cos\frac{2\pi l\rt{1}}{1+\rt{1}}\right )
\cos\left (2\pi lf_0 x\right )
\\ \nn
%\label{eq24}
&+&2\sum_{l\ge 1}\sum_{l'\ne l}\frac{(-1)^{l'-l}}
{4\pi^4(l'-l)^2}\frac{(1+\rt{1})^3}{\left (l+l'\rt{1}
\right )^2}
\\ \nn
&\times&\left (1-\cos\frac{2\pi(l'-l)\rt{1}}{1+\rt{1}}\right )
\cos\left (2\pi(l'f_0+lf_1)x\right ).
\ee

 \rf{eq22} is the main result for the
two-band problem at zero temperature. It contains the exact
amplitudes of the oscillating part of the energy. The oscillating
part of the magnetization is simply defined by
$m_{osc}=n_eG'(x)/2=\sum_F\ace(F)\sin(2\pi Fx)$. We notice that
$G(x)$ depends only on $\rt{1}$ and the fundamental frequencies
$\fa$. In the double sum \rfb{eq22}, it may happen that some terms
diverge because $l+l'\rt{1}$ vanishes in the denominator. But in
this case, the quantity $(l-l')\rt{1}/(1+\rt{1})=l$ is integer and
the interference term $1-\cos(2\pi l)$ vanishes at the same time.
It is easy to check that the quantity
$(1-\cos(\epsilon))/\epsilon^2$ is finite when $\epsilon$ tends to
0, and the divergences do not occur. The values of these singular
terms are given by this regularization. In the general case, we
can give the expression of the amplitudes $\ace(kF)$ for the
$k^{th}$ harmonics of $F=l_0f_0+l_1f_1$, where $l_1$ is a positive
integer and $l_0$ can be negative. If $f_0/f_1$ is equal to an
irreducible ratio of 2 integers $p_1/p_0$, then we can find many
combinations $l'f_0+lf_1$ in the double sum of \rf{eq22} that are
equal to $kF$. This occurs for $l=kl_1+mp_1$ and $l'=kl_0-mp_0$.
We obtain

\be\nn
\ace(kF)&=&\frac{kF(1+\rt{1})^3}{2\pi^3}
\\ \nn
&\times&
\sum_{m\ge 0}
\frac{(-1)^{k(l_0-l_1)-m(p_0+p_1)+1}}{(k(l_1+\rt{1}l_0)+m(p_1-\rt{1}p_0))^2}
\\ \label{eq33}
&\times &\frac{1-\cos \frac{2\pi\left (k(l_0-l_1)
-m(p_0+p_1)\right )\rt{1}}{1+\rt{1}}}{(k(l_0-l_1)-m(p_0+p_1))^2}.
\ee

In the case where $f_0/f_1$ is irrational, only the term $m=0$ has
to be taken into account:

\be \label{eq+}
\ace(kF)&=&\frac{(-1)^{k(l_0-l_1)+1}(1+c_1)^3F}
{2\pi^3k^3(l_0-l_1)^2(l_1+c_1l_0)^2}
\\ \nn
&\times& [1-\cos\frac{2\pi k(l_0-l_1)c_1}{1+c_1}]. 
\ee

The amplitudes \rf{eq33} holds for any
frequency except for the special case $k(f_0+f_1)$. Indeed
$\ace(k(f_0+f_1))$ is obtained with the condition $p_0\ne p_1$:

\be\nn
& &\ace(k(f_0+f_1))=\frac{\rt{1}}{1+\rt{1}}\frac{1}{\pi k}+
\\ \nn
& &\frac{k}{2\pi^3(p_0+p_1)^2}
\sum_{m\ge 1}\frac{(1+\rt{1})^3
(-1)^{m(p_0+p_1)+1}}{m^2(k(1+\rt{1})+m(p_1-\rt{1}p_0)
)^2}
\\ \label{eq26}
&\times&\left (
1-\cos\frac{2\pi m(p_0+p_1)\rt{1}}{1+\rt{1}} \right ).
\ee

In the case where $f_0/f_1$ is irrational, only the first term on
the right hand side of \rf{eq26} has to be taken into account, and
we obtain a simple result: the $k$ dependence is the same as it
would be in GCE case for a single band with fundamental frequency
$f_0+f_1$ (with the exception of a $(-1)^k$ term that appear in
the GCE case):

\be\label{eq27}
\ace(k(f_0+f_1))=\frac{\rt{1}}{1+\rt{1}}\frac{1}{\pi k}.
\ee

Other interesting cases are harmonics of $f_1$ and $f_0$.
The amplitudes for $kf_1$ are obtained by considering
\rf{eq22} with $l=k+mp_1$ and $l'=-mp_0$:

\be\nn
\ace(kf_1)&=&\sum_{m\ge 0}
\frac{(1+\rt{1})^3(-1)^{k+m(p_0+p_1)+1}}{(k+m(p_0+p_1))^2
(k+m(p_1-\rt{1}p_0))^2}
\\ \label{eq28}
&\times&\left (1-\cos\frac{2\pi
(k+m(p_0+p_1))\rt{1}}{1+\rt{1}}
\right )\frac{kf_1}{2\pi^3}.
\ee

In the irrational case, only the first term $m=0$ contributes:

\be\label{eq29}
\ace(kf_1)=\frac{(-1)^{k+1}(1+\rt{1})^3f_1}{2\pi^3k^3}\left (1-\cos\frac{2\pi
k\rt{1}}{1+\rt{1}}
\right ).
\ee

For the harmonics $kf_0$, when $f_0/f_1$ is rational, we
take $l=mp_1$ and $l'=k-mp_0$, with the condition $k\ne m(p_0+p_1)$,
and we obtain

\be\nn
\ace(kf_0)&=&\sum_{m\ge 0}
\frac{(1+\rt{1})^3(-1)^{k-m(p_0+p_1)+1}}{(k-m(p_0+p_1))^2
(k\rt{1}+m(p_1-\rt{1}p_0))^2}
\\ \label{eq30}
&\times&\left (1-\cos\frac{2\pi(k-m(p_0+p_1))\rt{1}}{1+\rt{1}}\right )
\frac{kf_0}{2\pi^3}.
\ee

If $f_0/f_1$ is not a rational, only the first term $m=0$ contributes
as before:

\be\label{eq31}
\ace(kf_0)=\frac{(-1)^{k+1}(1+\rt{1})^3f_0}{2\pi^3k^3\rt{1}^2}
\left (1-\cos\frac{2\pi k\rt{1}}{1+\rt{1}}\right ).
\ee

 Usually the terms $m\ge 1$ are only small corrections
due to the $1/m^4$ behavior. Finally, the ``forbidden'' frequency
amplitude corresponding to $f_1-f_0$ is given by taking $l=1+mp_1$
$l'=-1-mp_0$ in \rf{eq22}

\be\label{eq32}
\ace(f_1-f_0)=\sum_{m\ge 0}
\frac{(1+\rt{1})^3}{(2+m(p_0+p_1))^2
}
\\ \nn
\times\frac{(-1)^{m(p_0+p_1)+1}}{(1-\rt{1}+m(p_1-\rt{1}p_0))^2}
\\ \nn
\times\left (1-\cos\frac{2\pi(2+m(p_0+p_1))\rt{1}}{1+\rt{1}}\right )
\frac{f_1-f_0}{2\pi^3}.
\ee

For the irrational case, we just have

\be\label{eq34}
\ace(f_1-f_0)=\frac{(1+\rt{1})^3}{(1-\rt{1})^2}\frac{f_0-f_1}{(2\pi)^3}
\left (1-\cos\frac{4\pi\rt{1}}{1+\rt{1}}\right ).
\ee

These results imply that the jumps of the magnetization at zero temperature
only come from the contribution of the $f_0+f_1$ harmonics in the
case when $f_0/f_1$ is irrational (or $p_0$, $p_1$ infinite).
Indeed, a jump in magnetization means that the absolute value
of the $k^{th}$ harmonics in the free
energy or grand potential Fourier spectrum decreases for large $k$
like $1/k^2$ (respectively $1/k$ for magnetization).
Only in \rf{eq22} and \rf{eq27} this condition is satisfied. In other
cases \rfb{eq33}, the decreasing is only like $1/k^4$ ($1/k^3$),
as for the frequencies corresponding to $kf_0$ \rfb{eq31} or $kf_1$ \rfb{eq29}.

\begin{figure}                                               % Fig 4
\centering
\resizebox{\columnwidth}{!}{\includegraphics*{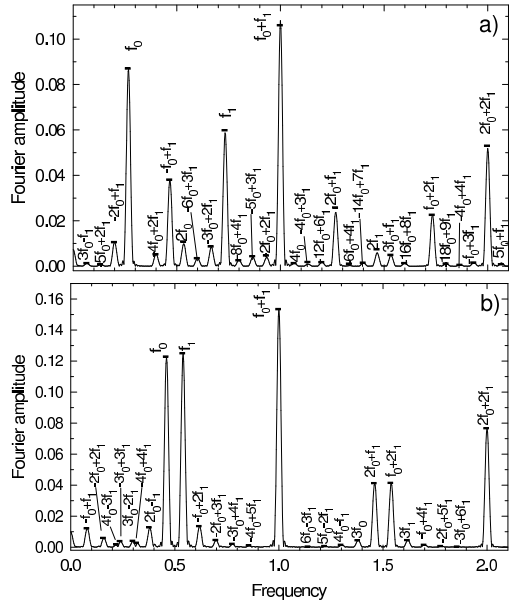}}
\caption{\label{TF_CE} Fourier analysis of dHvA spectra at zero
temperature computed in the canonical ensemble case. Horizontal
dashes are results deduced from Eqs. \ref{eq+}, \ref{eq27},
\ref{eq29}, \ref{eq30} and \ref{eq31}. The band parameters are
$\Delta_0 - \Delta_1$ = 1/10, c$_0^{-1}$ = 1, c$_1^{-1}$ = 2
(leading to f$_0$ = 4/15 and f$_1$ = 11/15) and $\Delta_0
-\Delta_1$ = 2/47, c$_0^{-1}$ = 1, c$_1^{-1}$ = 2670/2867 (leading
to f$_0$ = 52/113 and f$_1$ = 61/113) in Figs. \ref{TF_CE}a and
\ref{TF_CE}b, respectively.}
\end{figure}

Fig. \ref{TF_CE} compare the Fourier analysis of the dHvA spectrum
deduced from numerical computation of the chemical potential [see
Eq. (\ref{eq3})] to the predictions of Eqs. (\ref{eq+}),
(\ref{eq27}) and (\ref{eq29}) to (\ref{eq31}) (see horizontal
dashes in Fig. \ref{TF_CE}). An excellent agreement between
numerical computations and analytical formulae is obtained even in
the case where a given frequency results from several
combinations. This is e. g. the case of the Fourier peak at F =
8/15  in Fig. \ref{TF_CE}a which results not only from $2f_0$ but
also from e. g. $13f_0-4f_1$ and $-9f_0+4f_1$. Nevertheless, these
two other Fourier components have a small amplitude when compared
to $A_{CE}(2f_0)$ so that formulae such as  \rf{eq30} which is
relevant for this example, remain a good approximation in that
case.

%\begin{figure}                                               % Fig 1
%\centering
%\resizebox{\columnwidth}{!}{\includegraphics[angle=270]{amplitudes_CE.eps}}
%\caption{\label{Temp_CE} Absolute Fourier amplitudes of dHvA spectra
%as function of temperature as well as zero temperature components
%deduced from \rf{eq26} to \rf{eq34}. The band parameters
%are  $\Delta_0 - \Delta_1$ = 1/10, c$_0^{-1}$ = 1, c$_1^{-1}$ = 2
%(leading to f$_0$ = 4/11 and f$_1$ = 7/11, i.e. $p_0=11$ and $p_1=4$).
%Note that $A(2f_1)$ vanishes at one temperature and change sign.}
%\end{figure}

\section{Conclusion}

The Fourier components of the de Haas-van Alphen oscillation spectrum
have been calculated for an idealized two-dimensional Fermi liquid
with two parabolic bands in the case of canonical and grand
canonical ensembles. Exact analytical calculations of the Fourier
components at zero temperature for the CE [see Eqs. (\ref{eq+}),
(\ref{eq27}) and (\ref{eq29}) to (\ref{eq31})] are in excellent
agreement with numerical computations of the chemical potential
\rfb{eq3}, as demonstrated in Fig. \ref{TF_CE}. Regarding the high
$t/b$ range, asymptotic formulae have been derived for the CE and
GCE [see Eqs. (\ref{eq7.3}) to (\ref{eq7.6})]. Addition rule of
the effective masses associated to the combination frequencies
observed in the CE has been derived (see \rf{7.7}). Nevertheless,
and contrary to the GCE case, the $t/b$ dependence of the Fourier
components' pre-exponential term is non-trivial. In particular,
the 2$^{nd}$ harmonics amplitude can vanishes and changes sign as
$t/b$ varies. A good agreement is obtained between analytical
formulae and numerical computations for $\temp/b$ higher than 0.15
in the less favorable cases (see Figs. \ref{A(tx)_CE} to
\ref{Moscill}). Such a value which corresponds to e. g.
$T/B\gtrsim0.06$ $K/T$ for $\kappa$-(ET)$_2$Cu(SCN)$_2$, is
suitable for the determination of the effective masses from
experimental data with a good accuracy in a large temperature and
magnetic field range.

\appendix
\setcounter{equation}{0}
\section{}\label{app.1}

The series $\sum_{l=-\infty}^{\infty}\exp(-i\omega l)m_0(l)$
can be evaluated by considering the following periodic function:

\be\nn
\phi(x)=\mm_0(x)-\frac{\rt{1}}{1+\rt{1}}x,
\ee

with

\be\nn
\mm_0(x)=\left  [\frac{\rt{1}}{1+\rt{1}}(x+\delta)+
\frac{1}{2}\right ].
\ee

It is indeed simpler for calculations to consider $\mm_0$ instead of
$m_0$, both functions having the same values for integers
$\mm_0(l)=m_0(l)$.
We have $\phi(x+p+q)=\phi(x)$, and therefore the Fourier transformation
is written as

\be\label{eq1.1}
\phi(x)&=&\frac{\rt{1}\delta}{1+\rt{1}}
\\ \nn
&+&\sum_{l\ge 1}\frac{(-1)^l}{\pi l}
\sin\left (2\pi l\frac{\rt{1}}{1+\rt{1}}(x+\delta)\right ).
\ee

From that result, we have

\be\nn
\sum_{l=-\infty}^{\infty}\exp(-i\omega l)m_0(l)=
\sum_{l=-\infty}^{\infty}\exp(-i\omega l)\left (
\phi(l)+\frac{\rt{1}}{1+\rt{1}}l\right ),
\ee

the second sum on the right hand side of the previous
equation only contributes to derivatives of $S$. The sum
over $\phi(l)$ can be performed if we use \rf{eq1.1} and
commute the double sum. We obtain \rf{eq20}:

\be
& &\sum_{l=-\infty}^{\infty}\exp(-i\omega l)
\phi(l)
=\frac{\rt{1}\delta}{1+\rt{1}}S(\omega)
\\ \nn
&+&\sum_{l\ne 0}\frac{(-1)^l\exp(2i\pi l\rt{1}\delta/(1+\rt{1}))}
{2i\pi l}S\left (\omega-\frac{2\pi l\rt{1}}{1+\rt{1}}\right ).
\ee

We use the same analysis to compute the series
$\sum_{l=-\infty}^{\infty}\exp(-i\omega l)m_0(l)^2$. Here
we have to consider instead the function

\be
\phi(x)=\left (\mm_0(x)-\frac{\rt{1}}{1+\rt{1}}x\right )^2.
\ee

This is also a periodic function of $x$ with period $p+q$,
and it can be Fourier transformed. As $\phi(l)\propto m_0(l)^2$,
we obtain after further computation \rf{eq21}. The absolute value in
\rf{eq15} is periodic with period $p+q$ and therefore the function
we can consider for example is

\be
\phi(x)=\left |
(1+\rt{1})\mm_0(x)-\rt{1}(x+\frac{1}{2})+\frac{1}{2}-\rt{1}\delta
\right |.
\ee

As previously it can be Fourier transformed and the series
$\sum_{l=-\infty}^{\infty}\exp(-i\omega l)\phi(l)$ is expressed
as a sum of functions $S(\omega)$. We obtain

\be
& &\sum_{l=-\infty}^{\infty}\exp(-i\omega l)\phi(l)=
\frac{1+\rt{1}^2}{2(1+\rt{1})}
\\ \nn
&+&\sum_{l\ne 0}
(1-\rt{1})
\frac{(-1)^l\exp(2i\pi l\delta\rt{1}/(1+\rt{1}))}
{2i\pi l}
\\ \nn
&\times&S(\omega-2\pi l
\frac{\rt{1}}{1+\rt{1}})
\\ \nn
&+&\sum_{l\ne 0}
\frac{(-1)^l\exp(2i\pi l\delta\rt{1}/(1+\rt{1}))}
{2\pi^2 l^2}
\\ \nn
&\times&\left [1-\exp(-2i\pi l\frac{1}{1+\rt{1}})\right ]
S(\omega-2\pi l
\frac{\rt{1}}{1+\rt{1}}).
\ee

%\section{}\label{app.2}

\end{document}